\documentclass[conference]{IEEEtran}
\IEEEoverridecommandlockouts
\usepackage{cite}
\usepackage{amsmath,amssymb,amsfonts}
\usepackage{graphicx}
\usepackage{textcomp}
\usepackage{xcolor}
\usepackage{soul}
\usepackage{subcaption}
\usepackage{footnote}
\usepackage{hyperref}
\usepackage{csquotes}
\usepackage{algpseudocode}
\usepackage{multirow}
\usepackage{multicol}

\def\BibTeX{{\rm B\kern-.05em{\sc i\kern-.025em b}\kern-.08em
T\kern-.1667em\lower.7ex\hbox{E}\kern-.125emX}}

\hypersetup{ 
    colorlinks=true,
    linkcolor=black, 
    citecolor=green, 
    urlcolor=blue
}

\begin{document}

\title{Pediatric Wrist Fracture Detection in X-rays via YOLOv10 Algorithm and Dual Label Assignment System}

\author{
    \IEEEauthorblockN{Ammar Ahmed\IEEEauthorrefmark{1}\thanks{Email: ammaa@stud.ntnu.no},
    Abdul Manaf\IEEEauthorrefmark{2}\thanks{Email: amanaf.bscsf19@iba-suk.edu.pk}}
    \IEEEauthorblockA{
    \begin{tabular}{c}
        \IEEEauthorrefmark{1}Department of Computer Science (IDI), Norwegian University of Science \& Technology (NTNU), Gjøvik, 2815, Norway \\
        \IEEEauthorrefmark{2}Department of Computer Science, Sukkur IBA University, Sukkur, 65200, Pakistan \\
    \end{tabular}
    }
}

\maketitle

\begin{abstract}
Wrist fractures are highly prevalent among children and can significantly impact their daily activities, such as attending school, participating in sports, and performing basic self-care tasks. If not treated properly, these fractures can result in chronic pain, reduced wrist functionality, and other long-term complications. Recently, advancements in object detection have shown promise in enhancing fracture detection, with systems achieving accuracy comparable to, or even surpassing, that of human radiologists. The YOLO series, in particular, has demonstrated notable success in this domain. This study is the first to provide a thorough evaluation of various YOLOv10 variants to assess their performance in detecting pediatric wrist fractures using the GRAZPEDWRI-DX dataset. It investigates how changes in model complexity, scaling the architecture, and implementing a dual-label assignment strategy can enhance detection performance. Experimental results indicate that our trained model achieved mean average precision (mAP@50-95) of 51.9\%  surpassing the current YOLOv9 benchmark of 43.3\% on this dataset. This represents an improvement of 8.6\%. The implementation code is publicly available at https://github.com/ammarlodhi255/YOLOv10-Fracture-Detection
\end{abstract}


\section{\textbf{Introduction}}
Wrist fractures are among the most common pediatric injuries, leading to approximately 500,000 emergency department visits annually in the UK alone \cite{evidently2022wrist}. This high frequency imposes a significant burden on healthcare systems. Since children’s bones are still developing, fractures pose a risk of damaging growth plates. This is particularly concerning for wrist fractures, where damage to the distal radius growth plate can result in stunted growth and discrepancies in bone length \cite{erickson2024wrist}. Initial X-rays may not always reveal these fractures clearly, which can lead to delayed or missed diagnoses if not thoroughly evaluated by specialists \cite{wrist_fractures2024}. Additionally, managing pediatric wrist fractures requires substantial healthcare resources, including emergency care, imaging, follow-up visits, and potentially surgical interventions \cite{evidently2022wrist}. In response to these challenges, there has been growing interest in automating fracture detection in trauma X-rays. Recent studies have shown that AI systems can achieve high accuracy in detecting fractures, sometimes matching or even surpassing the performance of human radiologists \cite{gleamer2024, hussain2023bone, ahmed2024enhancing}. YOLO (You Only Look Once) object detection algorithms, in particular, have demonstrated significant advancements in this field \cite{ahmed2024enhancing, chien2024yolov8, ju2023fracture, nagy2022pediatric}. 

Recent advancements with YOLOv8 and YOLOv9 have shown these models outperforming their predecessors \cite{ahmed2024enhancing, chien2024yolov8, ju2023fracture, chien2024yolov9}. The high accuracy and speed of YOLO models position them as valuable tools for supporting clinical decisions in fracture diagnosis \cite{samothai2022evaluation}. In this study, we provide a first comprehensive analysis of YOLOv10 \cite{yolov10} and its variants using a large annotated GRAZPEDWRI-DX dataset \cite{nagy2022pediatric}.

The main contributions of this paper are as follows:
\begin{enumerate}
\item{We thoroughly evaluate all YOLOv10 variants on the GRAZPEDWRI-DX dataset, marking the first application of YOLOv10 to this dataset.}
\item{We achieve state-of-the-art performance with the mAP@50-95 metric on the GRAZPEDWRI-DX dataset, surpassing the results previously achieved with YOLOv9.}
\item{We explore the relationship between the complexity of the YOLOv10 model and its performance in fracture detection.}
\end{enumerate}

\begin{figure*}
\centering
\includegraphics[width=0.8\textwidth, height=5.6cm]{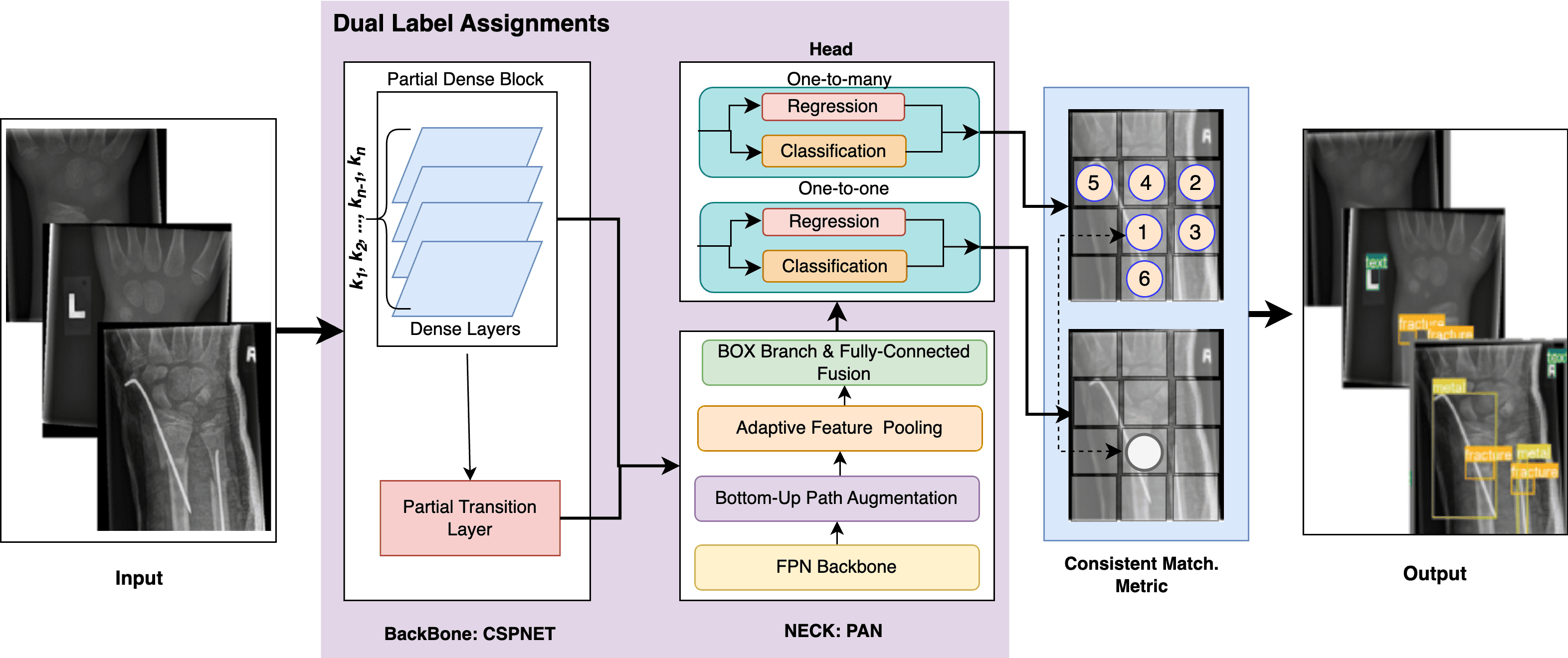}
\caption{YOLOv10 Architecture depicting the input, backbone, neck, head, and output. }
\label{fig1:yolov10}
\end{figure*}
\section{\textbf{Related Work}}
The YOLO series has gained extensive application in fracture detection due to its rapid inference speed and success across various detection tasks. Sha et al. \cite{Sha_Wu_Yu_2020a} employed YOLOv2 to detect fractures in a dataset of 5,134 spinal CT images, achieving a mean average precision (mAP) of 0.75. Hr{\v{z}}i{c} et al. \cite{hrzic2022fracture} compared YOLOv4’s performance against U-Net segmentation and radiologists on the GRAZPEDWRI-DX dataset. Their study involved two YOLOv4 models: one for identifying the most probable fractured object and the other for counting fractures. The first model attained an AUC-ROC of 0.90 and an F1-score of 0.90, while the second model achieved an AUC-ROC of 0.90 and an F1-score of 0.96. Further advancements were made with YOLOv5, as applied by the authors of the GRAZPEDWRI-DX dataset \cite{nagy2022pediatric}, who achieved a fracture detection mAP of 0.93 and an overall mAP of 0.62 at an IoU threshold of 0.5 using the COCO pre-trained YOLOv5m variant. Ahmed et al. \cite{ahmed2024enhancing} expanded this work by testing YOLOv5, YOLOv6, YOLOv7, and YOLOv8 along with their various compound-scaled variants on the GRAZPEDWRI-DX dataset. They found that the YOLOv8x variant achieved the highest overall mAP of 0.77 and a fracture detection mAP of 0.95. Ju et al. \cite{ju2023fracture} used YOLOv8 for detecting fractures and other wrist pathologies in the same dataset, reporting an mAP of 0.64, recall of 0.59, and precision of 0.73. Chien et al. \cite{chien2024yolov9} extended this research by applying YOLOv9, achieving the highest mAP@50 score of 0.66 and mAP@50-95 score of 0.44. They also explored YOLOv8 with an attention mechanism, resulting in the YOLOv8-AM model based on ResBlock and CBAM, which attained an overall mAP score of 0.66 \cite{chien2024yolov8}. Dibo et al. \cite{dibo2024deeploc} enhanced performance on the GRAZPEDWRI-DX dataset by integrating the GAM attention mechanism with the SwinTransformer backbone in YOLOv7, achieving an overall mAP score of 0.65 and an average precision (AP) of 0.94 for the fracture class. Despite these advancements, YOLOv10 has not yet been applied for fracture detection on the GRAZPEDWRI-DX dataset. This study aims to fill this gap by evaluating YOLOv10 and its variants on this dataset.

\section{\textbf{Material \& Methods}}

\subsection{Dataset}
A quantitative (experimental) study was conducted using data from 10,643 wrist radiography studies of 6,091 unique patients collected by the Division of Pediatric Radiology, Department of Radiology, Medical University of Graz, Austria \cite{nagy2022pediatric}. The dataset contains nine distinct objects shown in Table \ref{tab:table0}. It shows the number of X-ray images containing each object. Since the authors of the dataset did not provide a split, we randomly partitioned the dataset into a training set of 15,245 images (75\%), a validation set of 4,066 images (20\%), and a testing set of 1,016 images (5\%).

\subsection{Method}
High computational requirements lead to increased latency and consumption of more power which can delay real-time detection and response. This is particularly problematic in critical applications such as medical diagnostics, where timely decisions are crucial. YOLOv10 eliminates the need for non-maximum suppression (NMS). Instead of NMS, a dual-label assignment system is employed during training, which integrates one-to-many and one-to-one head strategies. This modification reduces computational burden and latency, which is crucial for real-time detection applications, especially in a clinical setting.  The model becomes more efficient in handling overlapping detections, which is particularly useful for detecting small, closely located pathologies or fractures. The model can focus on refining its bounding box predictions during training, potentially leading to higher precision in detecting small and subtle abnormalities in wrist X-rays or MRIs. The pipeline of YOLOv10 is shown in Fig.\ref{fig1:yolov10}.

\subsubsection*{Compact Inverted Block (CIB) Structure}
The architecture incorporates a Compact Inverted Block (CIB) structure that leverages depthwise convolutions for spatial mixing and pointwise convolutions for channel mixing. This design enhances efficiency while maintaining performance, which is crucial for deployment on devices with limited computational resources, such as portable X-ray or MRI machines.
\begin{figure*}
\centering
\includegraphics[width=1\textwidth, height=4.5cm]{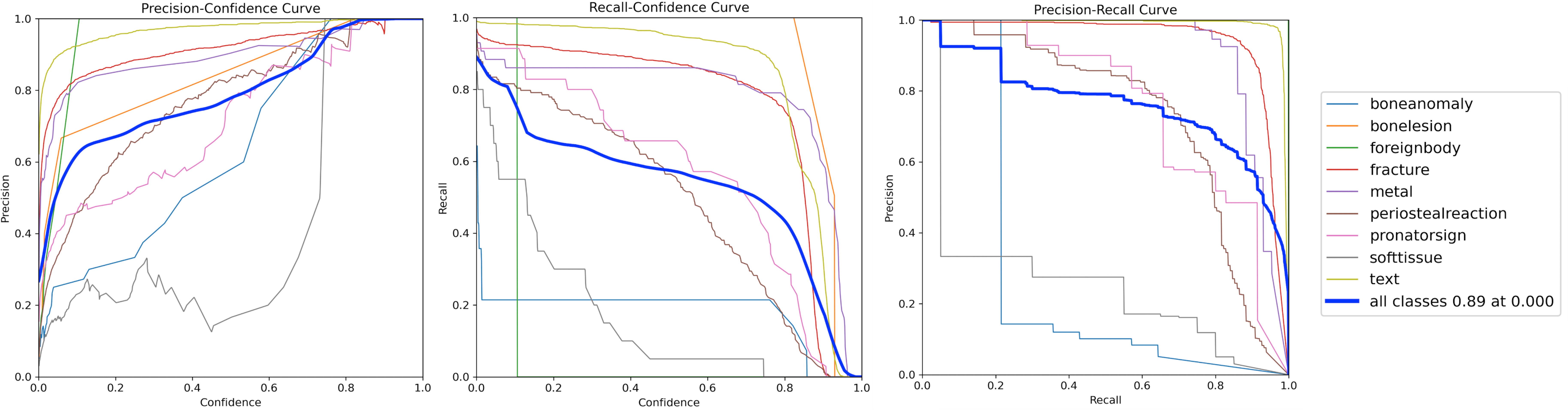}
\caption{Precision, recall, and PR curves of YOLOv10-M variant across increasing confidence scores. }
\label{fig1:curves}
\end{figure*}

\begin{figure}
\centering
\includegraphics[width=0.8\linewidth, height=5cm]{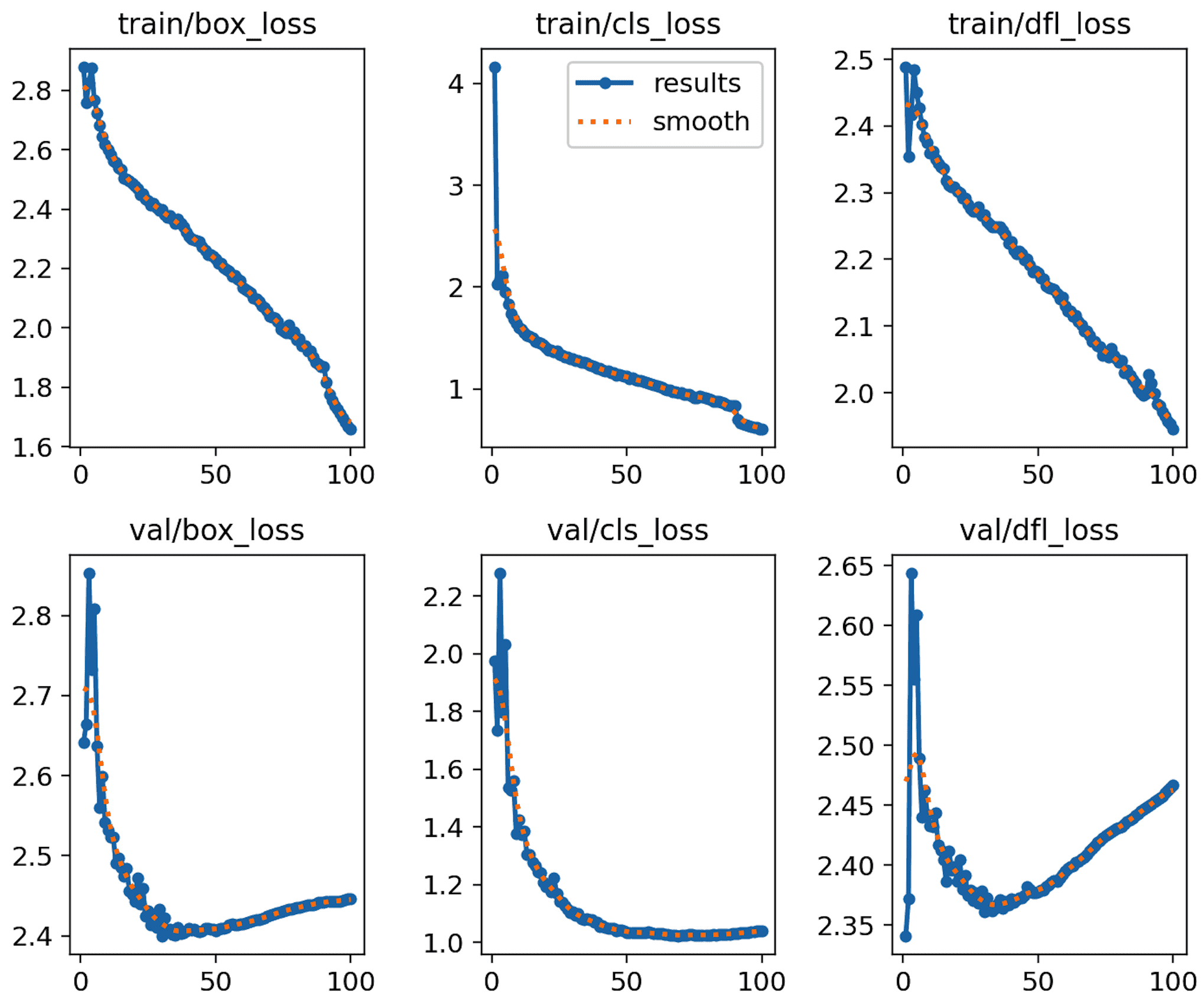}
\caption{Loss curves for YOLOv10-M variant. }
\label{fig1:results}
\end{figure}

\begin{table}[h]
\centering
\caption{Class Distribution}
    \begin{tabular}{l c c }
    \hline
    Abnormality & Instances & Ratio\\
    \hline 
  Boneanomaly                                                              & 192                                                               & 0.94\%                                                                            \\
  Bonelesion                                                                & 42                                                                & 0.21\%                                                                            \\
  Foreignbody                                                               & 8                                                                 & 0.04\%                                                                            \\
  Fracture                                                                 & 13550                                                             & 66.6\%                                                                           \\
  Metal                                                                     & 708                                                               & 3.48\%                                                                            \\
  Periostealreaction                                                        & 2235                                                              & 11.0\%                                                                           \\
  Pronatorsign                                                              & 566                                                               & 2.78\%                                                                            \\
  Softtissue                                                                & 439                                                               & 2.16\%                     \\
  
  Text                                                                & 20,274                                                               & 99.74\%                     \\
  \hline                                                      
  \end{tabular}
  \label{tab:table0}
\end{table}

\subsubsection*{Lightweight Classification Head}
YOLOv10 includes a decoupled classification head that separates the classification task from other tasks, such as bounding box regression, allowing each to be optimized independently. It employs depthwise separable convolutions, which factorize a standard convolution into a depthwise convolution followed by a pointwise convolution. This significantly reduces the number of parameters and computations.

\subsubsection*{Spatial-Channel Decoupled Downsampling}
This technique optimizes the processing of spatial and channel information, enhancing the detection of wrist pathologies and fractures. By decoupling spatial and channel downsampling, the method carefully manages information loss typically associated with downsampling, preserving critical features necessary for detecting small and subtle wrist pathologies. Spatial downsampling focuses on reducing dimensionality while maintaining key spatial structures, whereas channel transformation optimizes the mixing of feature information across channels.

\subsubsection*{Rank-Guided Block Design}
The Rank-Guided Block Design in YOLOv10 aims to improve model efficiency by addressing redundancy in the architecture. It calculates an intrinsic rank by analyzing the last convolution in the final basic block of each stage. The singular values of the convolutional filters are computed, and the intrinsic rank is determined by counting the number of singular values above a certain threshold. A lower intrinsic rank indicates higher redundancy, meaning many features do not contribute new information. A higher intrinsic rank signifies more diverse and informative features. This design ensures essential features are retained during optimization, enhancing the model’s ability to detect subtle and small wrist pathologies and fractures.

\begin{figure*}
\centering
\includegraphics[width=1\textwidth, height=8cm]{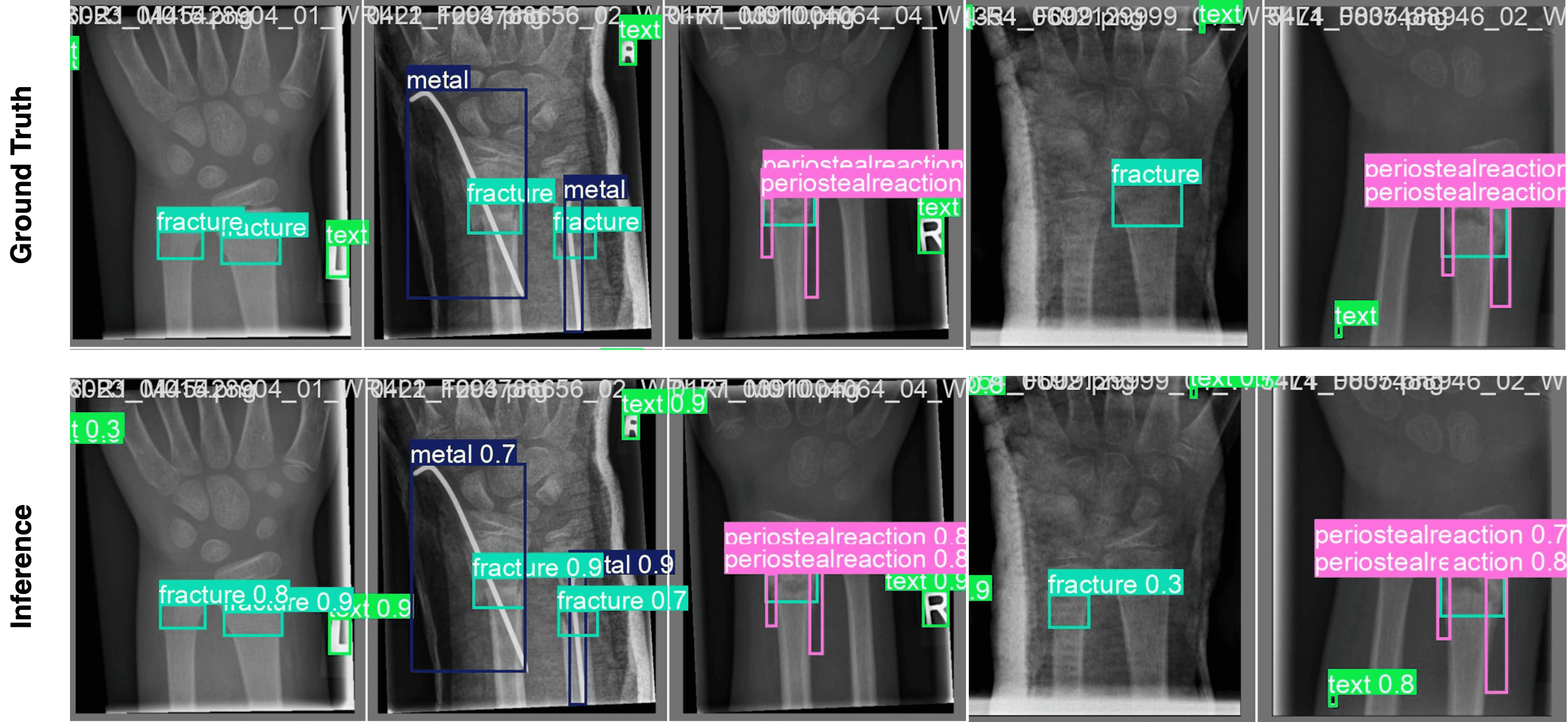}
\caption{Inferences made with YOLOv10-M variant. }
\label{fig1:inferences}
\end{figure*}

\subsubsection*{Large-Kernel Convolutions}

Large-kernel convolutions increase the receptive field of a model, which enhances its ability to capture information from larger objects. However, their indiscriminate use can lead to a loss of detail in smaller objects such as wrist pathologies. YOLOv10 mitigates these issues by employing large-kernel convolutions selectively: it uses 7×7 depthwise convolutions only in the deeper layers of the network, adds an extra 3×3 depthwise convolution branch during training for better optimization, and incorporates large kernels mainly in smaller model scales to avoid inefficiencies at larger scales.

\subsubsection*{Partial Self-Attention (PSA)} Self-attention mechanisms are effective for modeling global features but are often computationally expensive. YOLOv10 enhances efficiency with Partial Self-Attention (PSA) by partitioning features after a 1×1 convolution and applying self-attention to only part of these features. It efficiently fuses the processed features with an additional 1×1 convolution. It reduces the dimensions of the query and key vectors to half the size of the value vectors in Multi-Head Self-Attention (MHSA). PSA is strategically applied only after Stage 4 of the network, thereby minimizing computational overhead while still benefiting from global feature modeling.

\subsection{Experimental Setup}
The experiments in this study utilized NVIDIA A100-SXM4-80GB GPU, employing Python with the PyTorch framework. All YOLO variants were pre-trained on COCO before training on the GRAZPEDWRI-DX dataset. The models were trained for 100 epochs. Image resolution was set to 640 pixels, with a batch of 32. SGD optimizer was utilized with an initial and final learning rate (\(\eta = \eta_f = 1 \times 10^{-2}\)), a momentum (\(\mu = 9 \times 10^{-1}\)), and a weight decay (\(\wp = 5 \times 10^{-4}\)). During training, the ``best'' model was saved based on the highest validation (not test) accuracy achieved so far. 

\section{\textbf{Results \& Discussion}}
To evaluate the performance of YOLOv10 and YOLOv9 using our experimental design, this study compares several metrics: mean average precision at a 50\% IoU threshold (mAP@50), mean average precision across IoU thresholds from 50\% to 95\% (mAP@50-95), F1 score (the harmonic mean of sensitivity and precision), fracture sensitivity, number of parameters (Params), and floating-point operations per second (FLOPs).

\begin{table*}[h]
\centering
\caption{ Evaluation of YOLOv10 variants compared to YOLOv9 on all classes of GRAZPEDWRI-DX dataset. }
    \begin{tabular}{l c c c c c }
    \hline
    \textbf{Variant} & \textbf{mAP@50 (\%)} & \textbf{mAP@50-95 (\%)} & \textbf{F1 (\%)} & \textbf{Params (M)} &\textbf{ FLOPs (G)}  \\
    \hline 
     YOLOv9-C \cite{chien2024yolov9} & 65.3 & 42.7 & 64.0 & 51.0 & 239.0   \\
  YOLOv9-E \cite{chien2024yolov9} & 65.5 & 43.3 & 64.0 & 69.4 & 244.9   \\
    \hline
  YOLOv9-C' & 66.2 & 45.2 & 66.7 & 25.3 & 102.4   \\
  YOLOv9-E' & 67.0 & 44.9 & 70.9 & 57.4 & 189.2   \\
    \hline
  YOLOv10-N & 59.5 & 39.1 & 63.0 & 2.7 & 8.2\\
  YOLOv10-S & 76.1 & 51.7 & 67.5 & 8.0 & 24.5  \\
  YOLOv10-M & 75.9 & 51.9 & 69.2 & 16.5 & 63.5  \\
  YOLOv10-L & 70.9 & 46.6 & 68.7 & 25.7 &  126.4  \\
  YOLOv10-X & 76.2 & 48.2 & 69.8 & 31.6 &  169.9  \\
    \hline               
  \end{tabular}
  \label{tab:table1}
\end{table*}

\begin{table}[h]
\centering
\caption{ Evaluation of YOLOv10 variants on fracture class only. All values are shown in percentages (\%).}
    \begin{tabular}{l c c c}
    \hline
    \textbf{Variant} & \textbf{mAP@50} & \textbf{mAP@50-95} & \textbf{Sensitivity} \\
    \hline 
 YOLOv10-N & 93.8 & 56.9 & 88.8 \\
  YOLOv10-S & 93.7 & 56.9 &  88.1  \\
  YOLOv10-M & 94.0 & 58.4 & 92.5  \\
  YOLOv10-L & 94.9 & 58.8 & 89.1   \\
  YOLOv10-X & 94.0 & 57.8 & 91.4  \\
    \hline               
  \end{tabular}
  \label{tab:table2}
\end{table}

Table \ref{tab:table1} presents the evaluation results of the employed YOLO variants. The top two rows show the benchmarked YOLOv9 results on the GRAZPEDWRI-DX dataset \cite{chien2024yolov9}, while the third and fourth rows display the same YOLOv9 variants trained on our dataset split for a fair comparison. Thus, we highlight both the benchmarked YOLOv9 results and those using our experimental design. The remaining rows show the evaluation of YOLOv10 variants. Apart from the YOLOv10-N variant, all other YOLOv10 variants outperform the YOLOv9 variants across all metrics. Among the YOLOv10 variants, the X variant achieved the highest mAP@50 score of 76.2\%, while the M variant achieved the highest mAP@50-95 score of 51.9\%. Table \ref{tab:table2} shows the performance evaluation of YOLOv10 variants specifically for fracture pathology. The M variant performs exceptionally well, achieving a mAP@50 score of 94\% and the highest fracture sensitivity of 92.5\%. Therefore, we select the M variant for a thorough analysis, as it provides a balanced performance across all metrics.

The training and validation loss curves for the YOLOv10-M variant are illustrated in Fig.\ref{fig1:results}. Fig.\ref{fig1:curves} presents the precision, recall, and PR curves for the YOLOv10-M variant on the test set, both overall and for individual classes. Additionally, Fig.\ref{fig1:inferences} displays some inferences made by this variant compared to the ground truth. The YOLOv10-M variant demonstrates the ability to detect fractures with high confidence and efficiently identifies various pathologies, as evidenced by the second, more challenging instance. However, it should be noted that, although rare, the model can occasionally miss a fracture, as shown in the fourth image.

As shown in Table \ref{tab:table1}, increasing model complexity generally improves overall performance. However, while transitioning from the Nano variant (N) to the Small (S) and then to the Medium (M) variant significantly increased the mAP@50-95 score, further increases in complexity beyond the M variant led to a decrease in this score. The Large (L) and Extra-Large (X) variants may have introduced redundant parameters or layers without substantially enhancing pathology detection capabilities. The variant M of YOLOv10 achieves an optimal performance across all metrics on all classes including fracture. The model performs fewer floating point operations and has fewer parameters than its predecessors while surpassing them in performance.

\section{\textbf{Conclusion}}
In this study, we demonstrated the superiority of YOLOv10 variants over YOLOv9, with the YOLOv10-M variant achieving optimal performance across all classes, including fractures. Additionally, we showed that increasing the model complexity beyond this variant resulted in a decreased mAP@50-95 score. This study establishes a benchmark for YOLOv10 on the GRAZPEDWRI-DX dataset, providing a foundation for future research and development.

\section*{Acknowledgement}
This work was supported by the Curricula Development and Capacity Building in Applied Computer Science for Pakistani Higher Education Institutions (CONNECT) Project NORPART-2021/10502, funded by the Norwegian Directorate for Higher Education and Skills (DIKU).

\bibliographystyle{IEEEtran}
\bibliography{cas-refs}

\color{red}

\end{document}